\pdfoutput=1

\documentclass[11pt]{article}

\usepackage[]{ACL2023}

\usepackage{times}
\usepackage{latexsym}

\usepackage[T1]{fontenc}

\usepackage[utf8]{inputenc}

\usepackage{microtype}

\usepackage{inconsolata}

\usepackage{graphicx}
\usepackage{booktabs}
\usepackage{amsmath}
\title{Transparent and Scrutable Recommendations Using \\ Natural Language User Profiles}

\author{Jerome Ramos\textsuperscript{$\dagger$}
Hossein A. Rahmani\textsuperscript{$\dagger$}
Xi Wang\textsuperscript{$\dagger$}\textsuperscript{$\ddagger$}
Xiao Fu\textsuperscript{$\dagger$} 
Aldo Lipani\textsuperscript{$\dagger$} \\
        \textsuperscript{$\dagger$}University College London \\
        \textsuperscript{$\ddagger$}The University of Sheffield \\
        \texttt{\{jerome.ramos.20, hossein.rahmani.22, xiao.fu.20, aldo.lipani\}@ucl.ac.uk} \\
        \texttt{xi.wang@sheffield.ac.uk}}

\begin{document}
\maketitle
\begin{abstract}
Recent state-of-the-art recommender systems predominantly rely on either implicit or explicit feedback from users to suggest new items. While effective in recommending novel options, many recommender systems often use uninterpretable embeddings to represent user preferences. This lack of transparency not only limits user understanding of why certain items are suggested but also reduces the user's ability to scrutinize and modify their preferences, thereby affecting their ability to receive a list of preferred recommendations. Given the recent advances in Large Language Models (LLMs), we investigate how a properly crafted prompt can be used to summarize a user's preferences from past reviews and recommend items based only on language-based preferences. In particular, we study how LLMs can be prompted to generate a natural language (NL) user profile that holistically describe a user's preferences. These NL profiles can then be leveraged to fine-tune a LLM using only NL profiles to make transparent and scrutable recommendations. Furthermore, we validate the scrutability of our user profile-based recommender by investigating the impact on recommendation changes after editing NL user profiles. According to our evaluations of the model's rating prediction performance on two benchmarking rating prediction datasets, we observe that this novel approach maintains a performance level on par with established recommender systems in a warm-start setting. With a systematic analysis into the effect of updating user profiles and system prompts, we show the advantage of our approach in easier adjustment of user preferences and a greater autonomy over users' received recommendations.

\end{abstract}

\section{Introduction}
\label{sec:introduction}
\begin{figure}[h]
  \centering
  \includegraphics[width=\linewidth]{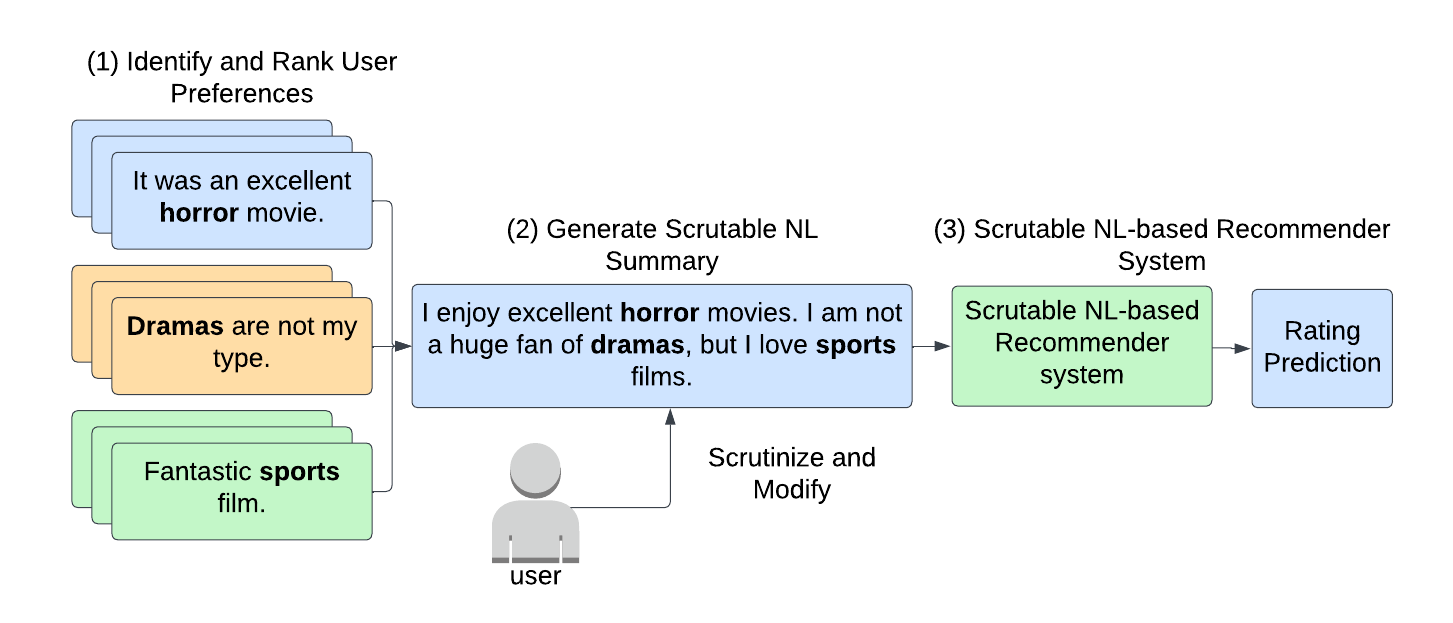}
  \caption{Overall architecture of UPR. In Step 1, we identify and rank user preferences from user reviews. Features are highlighted in \textbf{bold}. In Step 2, we use an LLM to generate a personalized, scrutable natural language (NL) profile based on the user's top features. In Step 3, we train a scrutable, NL-based recommender system using the generated NL profile to predict the user's rating for a target item.}
  \label{fig:architecture}
\end{figure}

Personalized recommender systems often rely on building latent representations from past user interactions to provide recommendations. These models, while effective, suffer from a lack of interpretability on the learned users' preferences, making it challenging to provide intuitive explanations for recommendations.  For example, many Collaborative Filtering (CF) techniques learn latent embeddings~\cite{he2017neural,pmf} to represent information about every user and item. These embeddings are complex, continuous vectors that cannot be easily interpreted or modified~\cite{user_profile_transparent_rec}. In fact, since collaborative filtering builds embeddings via past interactions, users would need to significantly change their interaction history to receive new recommendations that align with their current preferences. 

Given the strong performance of Large Language Models (LLMs) in a wide variety of NLP tasks~\cite{touvron2023llama, jiang2023mistral} as well as the inherent scrutable nature of text-based inputs, we explore how we can prompt LLMs with a natural language (NL) description of a user's preferences (i.e., user profiles) to provide greater transparency on how a model provides new recommendations. Furthermore, by using NL preferences rather than embeddings, a user can easily scrutinize and update their preferences when needed.

For recommender systems, we define \textit{transparency} as a faithful representation behind the recommendation mechanism and \textit{scrutability} as a direct and meaningful way to inspect and modify a user representation~\cite{explainable_rec}. Addressing these issues is vital: if users receive poor recommendations, they should understand why and have the means to modify their representation for better results. Research has shown that explanations that justify a recommendation help users make better and faster decisions and promote trust in the system~\cite{explainable_rec,tintarev_explain_rec}.

Recent works in transparent recommendation systems have explored using pre-defined templates~\cite{CIKM20-NETE} and automatically generated text~\cite{li-etal-2021-personalized,li2023personalized,Xie2022FactualAI} to explain recommendations to users. However, these works focused on explaining individual item recommendations rather than holistic, personalized preferences. Another approach to transparent and scrutable recommender systems is set-based preferences, which aim to explain a user's \textit{general preferences} for a particular domain~\cite{user_profile_transparent_rec,transparent_rec}. Specifically, rather than giving a rating for each individual item, a user can categorize groups of items into \textit{sets}, which are then used to recommend new items. For example, a user can state that they generally like `action' movies and generally dislike `romance' movies. This transparency lets users understand the model's interpretation and adjust their profile accordingly, leading to more aligned recommendations. Chang et al.~\cite{set_preferences} have shown that set-based preferences allow users to elicit preferences faster, leading to higher satisfaction with the resulting recommended items. ~\citet{sanner2023large} found that LLMs are competitive recommender systems in a near cold-start setting. In particular, they collected language-based user profiles from crowdsourced workers and found that by using few-shot prompting with the PaLM model~\cite{chowdhery2022palm}, they were able to achieve similar performance to item-based collaborative filtering methods.

In this paper, we propose \textbf{U}ser \textbf{P}rofile \textbf{R}ecommendation (UPR) depicted in Fig.~\ref{fig:architecture}, a language-based approach to set-based, transparent recommendations in a warm-start setting. Unlike a cold-start setting, a warm-start setting contains significantly more user-item interactions and could further improve the performance of collaborative filtering methods~\cite{he2017neural}. In lieu of a real-world dataset of NL profiles, which is both expensive and challenging to collect, we instead simulate users writing and modifying NL profiles by prompting instruction-tuned LLMs to generate profiles based on past user reviews. We first exemplify that these machine-generated NL profiles qualitatively summarize a user's preferences concisely while maintaining a scrutable format. Furthermore, according to the conducted experiments on two benchmark datasets, Amazon Review (Movies and TV) and TripAdvisor , we show how NL user profiles can be used to train a transparent, NL-based recommendation system based on scrutable NL profiles rather than uninterpretable user embeddings. We observe that our NL-based approach has competitive recommendation performance to popular baseline recommender systems while also being transparent and scrutable.

Our model enables a transparent and scrutinized recommendation process with competitive recommendation accuracy. In addition, past work suggests that a transparent and scrutable model is preferred because it allows users to understand the reasoning behind the model's recommendations and easily modify their preferences to receive new recommendations~\cite{user_profile_transparent_rec, transparent_rec}. To validate the advanced transparency and scrutability of our UPR model, we investigate how editing user profiles affects recommendation performance. For reproducibility, we publicly release all data and code\footnote{https://github.com/jeromeramos70/user-profile-recommendation}.

Our key contributions include:
\begin{itemize}
    \item A novel method to simulate NL user preference profiles from user feedback by prompting instruction-tuned LLMs, which are shown to be fluent, informative, concise, and relevant through a user study.
    \item A unique recommendation technique that prioritizes language-based input over learned embeddings, maintaining performance levels comparable to non-transparent baseline recommenders.
    \item An analysis of how editing a user profile impacts the downstream recommendation task.
\end{itemize}



\section{Related Works}
\label{sec:related_works}

\subsection{Explainable Recommendation}

Explainable recommendation aims to both generate relevant recommendations and provide a justification for why those items are being recommended. One of the two main perspectives of explainable recommendation research is explainable machine learning \cite{li2023personalized}. Explainable machine learning uses various computational techniques to understand and convey why a model returns a particular explanation. Examples of machine-learning-based explainable recommender systems include using pre-defined templates~\cite{CIKM20-NETE,transparent_rec}, counterfactual explanations~\cite{counterfactual_hiep_2021}, visualization~\cite{geng-etal-2022-improving}, and natural language generation for an item-level recommendation, meaning that the explanation is for a target user-item pair~\cite{li2023personalized, li-etal-2021-personalized}. Notably, these works~\cite{li2023personalized, li-etal-2021-personalized, CIKM20-NETE} often make use of user and item identifiers to create uninterpretable embeddings, meaning that a user would need to dramatically change their interaction history in order to update their recommendations. Rather than focusing on explaining a single recommendation (i.e., output) for a given user, we focus on making the \textit{input} of the model fully transparent and scrutable. By focusing on natural language as the input of the model, the user has full transparency on what data is used to generate recommendations and can seamlessly update the input as necessary. This approach can be seen as more similar to content filtering methods~\cite{books/sp/Aggarwal16, item-knn}, which matches the attributes of items with the interests or preferences indicated in a user's profile. However, by leveraging LLMs, we are able to use the pretrained knowledge that an LLM has on the item set and fine-tune a user's NL profile to increase performance on the recommendation task.

\subsection{Set-Based Preferences}

Many papers investigate the scenario where preferences are expressed over sets, which are then used to recommend individual items~\cite{Sharma_2019, transparent_rec, set_preferences}. These set-based preferences are then used to generate recommendations for individual items. One example of set-based preferences is tagging, where a user selects keyword(s) for each item. Examples of tags include ``superhero'', ``action'', etc. These tags help users indicate their preferences in their own personalized language. Similarly, ~\citet{mysoreEditableUserProfiles2023a} generates a transparent user profile by selecting a small set of human-readable concepts from a global inventory of keywords to enable controllable recommendations. Past research has shown that users are able to elicit preferences over sets faster than rating each item individually~\cite{set_preferences}. 

Most similar to our work is the transparent model presented by Sanner et al.~\cite{sanner2023large}, which studies how NL profiles can be used as input for LLMs in near-cold start settings to obtain competitive performance with baseline recommender models. The main novelty of our work is that we compare the performance of language-based preferences in a \textit{warm start} scenario, where recommender systems have access to significantly more user-item interactions. Furthermore, rather than collecting human written NL profiles like~\citet{sanner2023large}, we instead use LLMs to automatically generate an NL profile based on past reviews. ~\citet{mysoreLargeLanguageModel2023} has previously shown that historical user reviews can be used to generate synthetic narrative queries with LLMs for narrative-driven recommendation. Not only does this approach allow us to explore how LLMs can be used to summarize user preferences in a scrutable and editable format, but it also mitigates the time-intensive, costly task of crowdsourcing high-quality NL profiles written by users. These NL profiles serve as an automated way to generate set-based preferences from past reviews, which users can modify and edit as needed. Lastly, we are the first work that we are aware of that evaluates how NL profiles can be updated to receive new recommendations from an LLM-based recommender system in a warm-start setting.




\section{Methodology}
\label{sec:user_profile_generation}
\subsection{Goal of User Profiles}

Currently, there is no open-source, large-scale dataset of language-based preferences for recommendation. Furthermore, collecting such a dataset is both challenging and expensive because of the amount of data and quality needed to effectively fine-tune and benchmark a model. Instead, we leverage user text reviews and explicit ratings to infer a user's preferences and generate a corresponding NL profile. Not only do these reviews give us insight into which features the user likes and dislikes, but they also allow us to generate personalized profiles written with the user's own terminology. In addition, by generating NL profiles, users can automatically receive both their NL profiles and corresponding recommendations. The user can then adjust and modify the profile accordingly, which can save time versus manually crafting a prompt that provides accurate recommendations.

Balog et al.~\cite{transparent_rec} explored using tags, user-written keywords to describe an item, to generate template-based profiles. Although tags have been commonly used for recommendation tasks~\cite{DBLP:series/lncs/Bogers18,Chen2020JIT2RAJ}, past research has pointed out issues with tag quality~\cite{tag_quality}. In addition, tags may not provide sufficient granularity of preferences~\cite{user_profile_transparent_rec}. We argue that text-based profiles are closer to how a user naturally describes their personalized preferences versus a set of tags.


\subsection{Identifying and Ranking User Preferences}
\label{subsec:rank_pref}

Let $\mathcal{I}$ be the set of all items with respect to target user $u$. Each item $i \in \mathcal{I}$ has a (normalized) rating $r \in [-1, 1]$, where ratings lower than 0 mean the user disliked an item and ratings higher than 0 mean the user liked the item. Additionally, each item contains an explanation $e$, which is a sentence written by the user in their review that best describes their feedback for that item. Using a phrase-level sentiment analysis toolkit \cite{SIGIR14-Sentires}, we can also extract a feature word $f$ from $e$, where $f$ is the important keyword of the explanation. For example, in the explanation, ``the swimming \textbf{pool} is fantastic'', pool is the extracted feature. We can then infer a user's set-based preference for a particular feature $f$ by calculating the average rating of $\mathcal{I}_f$, the set of all items whose feature is $f$:
\begin{equation}
    \overline{r_f} = \sum_{i \in \mathcal{I}_f} r_i,
\end{equation}
where $\overline{r_f} < 0$ means that the user dislikes feature $f$ in general and $\overline{r_f} > 0$ means the user generally likes $f$. We then calculate a utility score $U(f)$ for a given feature $f$ to take into account how often a user uses a particular feature in reviews and subtract a discount statement based on its statistical strength. We define $U(f)$ as:
\begin{equation}
    U(f) =  |\overline{r}(f)| \times cov(f) \times sig(f),
\end{equation}
where $cov(f)$ and $sig(f)$ are the coverage and significance scores for a particular feature $f$. We define the coverage score as the ratio of items whose feature is $f$ over the entire item set.
\begin{equation}
    cov(f) = \frac{|\mathcal{I}_f|}{|\mathcal{I}|}.
\end{equation}
Additionally, we define the significance score as a discount statement where the statistical strength is less
than two standard errors, that is:
\begin{equation}
    sig(f) = min\left(2, |\overline{r_f}|/\frac{\sigma_t}{\sqrt{|\mathcal{I}_f|}}\right),
\end{equation}
where $\sigma_f$ is the variance of $R_f$, which is defined as the set of all ratings $r \in R$ with feature $f$. 
To reduce the noise in the ranking process, we grouped features by their stems, which were extracted using the NLTK Porter stemmer \cite{nltk}. By doing so, we can treat features such as ``pool'' and ``pools'' as a single entity rather than two separate features. We manually inspect the preprocessed data to ensure that the grouped entities are similar and remove features that are overly generic (e.g., the feature `film' in the movie domain).
Once the utility scores for the entire dataset have been calculated, we use the features with the highest scores to describe a user's preferences. Since we calculate significance scores using absolute values, features that are \textit{disliked} by the user may also be included in the profile.

Note that our user preference identification and ranking algorithm is similar to what was developed by Balog et al.~\cite{transparent_rec} for the MovieLens-20M (ML-20M) dataset~\cite{movielens}. However, there are a few key differences in our method, which mainly stem from how the user feedback data is structured. For example, our user review data contains only one feature per item rather than multiple tags per item. Thus, our coverage score rewards features that appear more in the dataset rather than adding a penalty if they are overused. In our scenario, we argue that this is beneficial because if a user writes a review with a certain feature often, it means that this feature is important to them.

\subsection{Transforming Preferences to NL Profiles}

After identifying a user's top features, we can transform the list of relevant reviews into an NL profile that captures a user's general preferences for the target domain. Using an instruction-tuned LLM, we can prompt the model to summarise the list of reviews to create a holistic profile that explains the user's general preferences for a particular domain (shown in Appendix~\ref{subsec:appendix-prof-gen}). After several iterations of prompt engineering, we provide 5 reviews per feature in the prompt, which fits within the max token constraint during experimentation. 
Since user-written reviews are used in the prompt, the resulting NL profile is generated with each user's personalized semantics. Since there is no ground truth for NL profiles, we instead analyze the quality of the profiles in Section~\ref{subsec:qualitative_study} and the utility of the generated profiles by evaluating the performance of a transparent recommender system trained on the NL profiles.

\subsection{Recommendation Task}

Similar to the P5 prompts for LLM-based recommendation~\cite{geng-etal-2022-improving}, we complete the rating prediction task by adding the NL profile to a prompt (shown in Appendix~\ref{subsec:appendix-rec-task}) and fine-tuning an LLM to complete a regression task. One key difference from traditional approaches is that no user identifiers or item identifiers are used during the NL profile generation or recommendation tasks. Thus, users can easily scrutinize the recommendation pipeline at any stage because all inputs are natural language-based rather than identifier-based. We further explore scrutability in Section~\ref{subsec:new_preferences}.


\section{Experimental Setup}
\label{sec:experimental_setup}
\subsection{Datasets}
We conduct our experiments on two publicly available explainable recommender datasets, Amazon Movies and TV (Amazon-MT)~\cite{He2016UpsAD} and TripAdvisor\footnote{\url{https://www.tripadvisor.com}} (hotels),  which focus on generating explanations for a user-item pair~\cite{CIKM20-NETE}. Each dataset consists of a user identifier, an item identifier, a rating from 1 to 5, and an explanation extracted from user reviews. We use an 8:1:1 train/validation/test split. The datasets were preprocessed to exclude any records where the title of the item is missing. To ensure a warm-start setting, every user and item appear at least once in each split, and there are at least 5 reviews per user in the training set. For reproducibility, we save all profiles generated for recommendation.
\begin{table}
  \centering
  \caption{Statistics of the Amazon-MT and TripAdvisor datasets.}
  \label{tab:freq}
  \resizebox{\columnwidth}{!}{%
    \begin{tabular}{lcc}
      \toprule
      & \textbf{Amazon-MT} & \textbf{TripAdvisor} \\
      \midrule
      \#users & 4,811 & 9,765 \\
      \#items & 5,459 & 6,280 \\
      \#records & 288,693 & 320,023 \\
      \#features & 5,114 & 5,069 \\
      \#records per user & 60.01 & 32.77 \\
      \#records per item & 52.88 & 50.96 \\
      \#words per explanation & 14.14 & 13.01 \\
      \bottomrule
    \end{tabular}%
  }
\end{table}

\subsection{Evaluation Metrics}

To measure recommendation accuracy, we use Root Mean Squared Error (\textbf{RMSE}) and Mean Absolute Error (\textbf{MAE}), two common metrics in the rating prediction task for these datasets~\cite{CIKM20-NETE, li2023personalized}. MAP with a relevance threshold of 4.0 and nDCG are calculated using condensed lists, following the evaluation strategy outlined by Sakai~\cite{sakai}; that is, we only consider items that are explicitly rated in the test set.

\subsection{Baselines}

We compare our model with the following baseline recommender systems. All baseline models are implemented using the Cornac library with default parameters~\cite{truong2021multi}.

\begin{itemize}
    \item \textbf{Most Popular}: An algorithm that recommends items with the most ratings.
    \item \textbf{UserKNN-BM25}~\cite{books/sp/Aggarwal16}: A collaborative filtering method that uses K-Nearest Neighbors to calculate distance between users with BM25 re-weighting.
    \item \textbf{Item-KNN-cosine}~\cite{item-knn}: A collaborative filtering method that uses K-Nearest Neighbors to calculate cosine distance between items.
    \item \textbf{MF}~\cite{2009korenmf}: Matrix Factorization learns latent user and item embeddings to provide recommendations.
    \item \textbf{NeuMF}~\cite{he2017neural}: a hybrid recommendation model that combines matrix factorization with multi-layer perceptrons.
\end{itemize}

In addition, we compare our model with two item-level explainable recommender models that describe why a user might like a target item. For example, given ``user A'' and ``item 123'' as input, the model might return a rating of 4.0 with the following explanation: ``the swimming pool is fantastic''. Note that both of these models are considered collaborative filtering, meaning that they mainly rely on user and item identifiers to generate recommendations. While not directly analogous due to our task of generating explainable NL profiles as input to a recommender model, this comparison yields insights into the performance of explainable models trained for rating prediction.
\begin{itemize}
    \item \textbf{PETER+}~\cite{li-etal-2021-personalized}: a small transformer designed for rating prediction and explanation generation using context prediction.
    \item \textbf{PEPLER-MLP}~\cite{li2023personalized} An item-level explainable model that fine-tunes GPT-2 for rating prediction and explanation generation for a user-item pair.
\end{itemize}


\subsection{Implementation Details}

For NL profile generation, we use the features defined in the datasets to identify and rank a user's set-based preferences. We then use a custom prompt for each user to generate an NL profile. We experiment with generating NL profiles using the 7B instruction-tuned versions of Llama2 \cite{touvron2023llama} and Mistral~\cite{jiang2023mistral}. We generate each NL profile with a maximum of 200 tokens, a number of features set to 5, and a temperature of 0.7. For reproducibility, we save all generated user profiles used to fine-tune our model\footnote{https://github.com/jeromeramos70/user-profile-recommendation}.

For the recommendation task, we train each model on the training set, tune hyperparameters on the validation set, and report the performance on the test set. The models are trained using either the Llama2 or Mistral-based profiles. We use GPT-2 \cite{radford2019language} as our pre-trained language model and train with a batch size of 8 and the AdamW optimizer \cite{2015-kingma} with a learning rate of 3e-4. For full implementation details, including prompts used and hyperparameter tuning, refer to Appendix~\ref{sec:appendix}.

\section{Results and Analysis}
\label{sec:results}
\subsection{Recommendation Performance}

\begin{table*}[!ht]
\centering
  \caption{Recommendation Performance on the Amazon-MT and TripAdvisor datasets. The best-performing values are highlighted in \textbf{bold}.}
  \label{tab:rec_performance}
  \begin{tabular}{lcccc|cccc}
    \toprule
    & \multicolumn{4}{c}{\textbf{Amazon-MT}} & \multicolumn{4}{c}{\textbf{TripAdvisor}}\\
    \midrule
    Model & RMSE & MAE & nDCG@10 & MAP & RMSE & MAE & nDCG@10 & MAP\\
    \midrule
    MostPop & 1.505 & 0.962  & 0.896  & 0.815  & 1.277  & 0.902  & 0.932  & 0.857  \\
    UserKNN & 0.960  & 0.707  & 0.935  &  0.864  & 0.836  & 0.634  & 0.952  & 0.883  \\
    Item-KNN & 1.045  & 0.790  & 0.897  & 0.823  & 0.890  & 0.683  & 0.953  & 0.852 \\
    MF & 0.925  & 0.686  & \textbf{0.941}  & \textbf{0.870}  & \textbf{0.786}  & \textbf{0.599}   & \textbf{0.960}  & \textbf{0.891} \\
    NeuMF & 0.943 & 0.694 & 0.936 & 0.866 & 0.819 & 0.634 & 0.955 & 0.886\\
    \hline
    PETER+ & \textbf{0.924}  & 0.685  & 
    \textbf{0.941}  & \textbf{0.870}  & 0.803  & 0.621  & 0.958  & 0.889 \\
    PEPLER-MLP & 0.925  & \textbf{0.672}  & \textbf{0.941}  & 0.869  & 0.793  & 0.606  & 0.959  & 0.889 \\
    \hline
    UPR (Llama2) & 0.944  & 0.678   & 0.938  & 0.866  & 0.804  & 0.616  & 0.955  & 0.885 \\
    UPR (Mistral) & 0.941  & 0.679   & 0.940  & \textbf{0.870}  & 0.804  & 0.610  & 0.952  & 0.887 \\
  \bottomrule
\end{tabular}
\end{table*}

We report the performance of our model compared to baseline models in Table \ref{tab:rec_performance}. We observe that UPR performs comparatively well with our baseline models and is consistently competitive across all metrics. Importantly, all compared models learn user preferences from \textit{all} interactions by taking advantage of the efficiency of embeddings. In contrast, UPR is limited to the information provided in the profile, which has a maximum length of 200 tokens. Consequently, UPR can only learn from a fraction of the interactions in the dataset. Nevertheless, UPR performs fairly well under the aforementioned constraints, meaning that LLMs can identify relevant items using only interaction data distilled into a short amount of text.

Even though the NL profile comparatively contains fewer features about the user, we observe that scrutable NL profiles can be used as a substitute for uninterpretable user embeddings without sacrificing much in performance. Given that newer models such as GPT-3 \cite{brown2020language} have shown better performance than GPT-2 \cite{radford2019language} and are trained on more data and natural language tasks, we anticipate that our model's performance will improve with larger-scale models. However, SOTA models contain significantly more parameters, meaning that more powerful hardware is needed to train them.

Given the competitive performance of UPR, we argue that an explainable model may be useful in real-life scenarios where human users value the transparency and scrutability of a set-based preference model. For example, if a user wanted to change their profile preferences in the collaborative filtering model, they would need to make many changes to their interaction history to see a substantial difference in their recommended items. For future work, we plan to study how human users interact with UPR and investigate the tradeoffs between traditional recommender methods and a transparent, scrutable recommender model.

\subsection{Effects of Number of Features on Recommendation Performance}

\label{subsec:prof_length}
\begin{figure}[!ht]
  \centering
  \includegraphics[width=\linewidth]{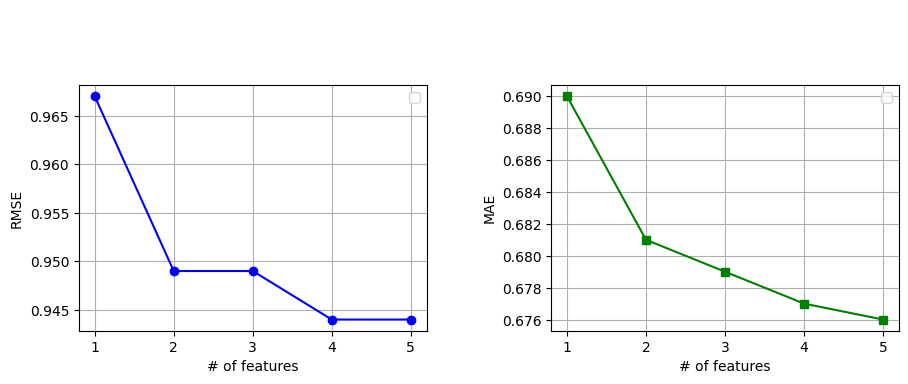}
  \caption{RMSE and MAE over a varying number of features in NL profile for the \textbf{Amazon-MT} dataset.}
  \label{fig:Amazon-MT}
\end{figure}

To examine the impact that the number of features contained in an NL profile has on recommendation performance, we run an ablation study that studies how changing $k$, the number of features used to generate the NL profile affects the downstream recommendation task. In particular, we prompt LLama2 to generate NL profiles but only use reviews from the top-k feature for each user, where $k \in \{1,2,3,4,5\}$.  Each set of NL profiles is then used to fine-tune UPR.

Observing the changes in RMSE and MAE in Figure~\ref{fig:Amazon-MT}, a discernible trend emerges: augmenting the count of features reduces the loss, but the improvement in performance sees diminishing returns as additional features are incorporated. This observed trend is rationalized by the arrangement of features in order of significance, implying that subsequently added features hold lesser relevance for the user. Furthermore, the diminishing returns of adding features provides empirical evidence that the feature ranking methodology in Section~\ref{subsec:rank_pref} helps select the best features to use in the profile.

An important point to note is that the incorporation of an excessive number of lower-ranked features may potentially introduce noise into the user profile. Given that a user profile can only accommodate a finite amount of text, this noise might, in turn, compromise the efficacy of the recommendations provided. As a follow-up study, we plan to research how to strategically select the number of features and reviews that best capture a user's holistic preferences and optimize the recommendation performance. Furthermore, we set the max token limit of an NL profile to 200 tokens. We found that this was qualitatively a reasonable max length for a profile. However, more research is needed to study the tradeoffs between the length of the profile with regard to recommendation performance and cognitive load for users. 

\subsection{Qualitative Case Study on NL Profiles}
\label{subsec:qualitative_study}
As there is no ground-truth data available for the NL profile generation task, we instead assess the quality of our generated NL profiles by conducting a qualitative case study with users. Five participants, all of whom are master's students at a university, were shown 50 NL profiles from each dataset and were tasked with answering whether each NL profile met each of the following criteria:
\begin{enumerate}
    \item \textbf{Fluency}: Is the NL profile both syntactically and semantically correct?
    \item \textbf{Informativeness}: Does the NL profile provide important information for a user profile?
    \item \textbf{Conciseness}: Is the NL profile written in a concise manner?
    \item \textbf{Relevance}: Given the list of reviews, is the NL profile relevant to the user?
\end{enumerate}

\begin{table}[!t]
  \centering
  \caption{User study measuring the quality of the generated NL profiles. We report the average number of satisfactory samples out of 50 samples per domain.}
  \label{tab:user_study}
  \begin{tabular}{l|cc}
      \toprule
       Metric & Amazon-MT & TripAdvisor\\
       \midrule
       Fluency & 95\% & 92\% \\
       Informativeness & 86\% & 82\% \\
       Conciseness & 75\% & 72\% \\
       Relevance & 90\% & 87\% \\
       \bottomrule
  \end{tabular}
\end{table}

Participants were shown positive and negative examples per criteria in order to improve annotation quality. Each question is answered with a ``yes'' or ``no''. We show the results of the user study in Table \ref{tab:user_study}. The Fleiss' kappa coefficient among annotators was 0.82, suggesting a very high level of agreement. Overall, annotators were highly satisfied with the quality of our generated NL profiles. Both Amazon-MT and TripAdvisor scored well on fluency, informativeness, and relevance but lower on conciseness. During follow-up discussions, annotators felt that the profiles contained few grammatical errors and were generally relevant and informative. However, they also pointed out that profiles could be excessively long at times and contain redundant information. Given that we used a zero-shot approach to generate profiles, we anticipate that NL profiles can be improved through either additional prompt engineering or fine-tuning on the NL profile generation task.

From manual inspection of the NL profiles for Amazon-MT, we noticed that Llama2 and Mistral were sometimes able to infer actors and characters of a film based on the review, even when not explicitly mentioned. Although this capability can potentially improve the downstream recommendation task, the ability to infer information can also lead to hallucinations. Consequently, the model can also generate inaccurate facts and negatively impact performance. Further research is needed to ensure that the NL profile generated is factual and aligns with the user's interests.


\subsection{Scrutinising NL Profile Preferences}
\label{subsec:new_preferences}

\begin{figure}[!t]
    \centering
    \includegraphics[width=\linewidth]{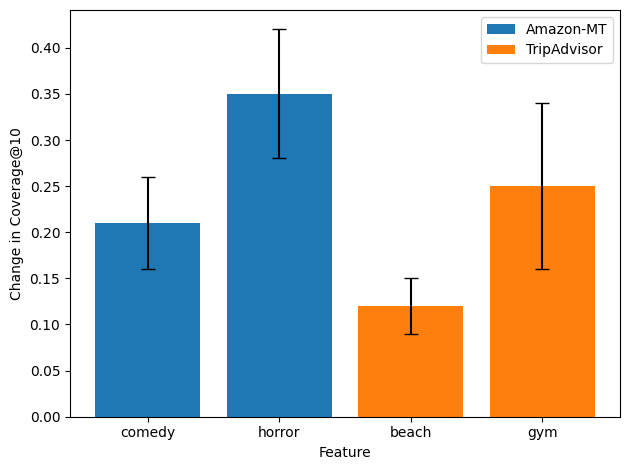}
    \caption{Change in Coverage@10 between edited profile versus original profiles when adding a new preference for a target feature.}
    \label{fig:counter-coverage}
\end{figure}

To evaluate the scrutability capabilities of UPR, we simulate a user adding a new preference to their NL profile. In particular, we select 200 random users from the original profiles $O$ that do not express any preference towards target feature $t$. We then prompt Mistral-Instruct-7B to edit the profile such that the user likes $t$ (shown in Appendix~\ref{subsec:appendix-editing-prof}) to obtain the edited profile set $C$. We randomly sample 100 items from the test set where feature $f=t$ and 100 items where $f\neq t$. Finally, we run the model at inference time and calculate the Coverage@10, which is defined as the number of items in the top-10 recommendations where $f=t$, divided by 10. We repeat this sampling process five times on both datasets and use different seeds for robustness. We report the change in Coverage@10 between $O$ and $C$ in Figure~\ref{fig:counter-coverage}, defined as $\overline{C_t} - \overline{O_t}$.

We observe that the Coverage@10 for all $t$ increases across all samples when using $C$, meaning that a user can simply add a preference towards $t$ to the NL profile, and the model will update the recommendations accordingly. In addition, the variance of the change in coverage shows that the recommendations can be correctly altered consistently.

Importantly, UPR does not require further fine-tuning when adding new preferences because the model learned how to interpret NL profiles effectively during the training phase. Furthermore, the features of each item were not explicitly passed to the model during fine-tuning or inference, meaning that the model has learned the features of each item from the NL profiles and its own pre-trained knowledge.

Overall, we show that \textit{UPR} is able to learn the features of each item during training, meaning that a user can scrutinize and edit their profile to instantly receive updated recommendations. In contrast, there is no simple way to update user preferences in traditional collaborative filtering methods. Consequently, a user would need to drastically change their interaction history to align the recommendations with their current interests.

It is important to note that the reviews used to determine the features during the sampling process may be written by an arbitrary user $A$ in Amazon-MT rather than the target user $B$. Thus, user $A$ might consider the item a `comedy' movie, but user $B$ may not use the same terminology because preferences are subjective. Furthermore, user $B$ still expresses preferences on other features. This means that they may highly rate non-target items in both $O$ and $C$, which can lead to a decreased change in coverage.


    


\section{Conclusion and Future Work}
\label{sec:conclusion}
In this work, we propose a novel recommendation model that replaces uninterpretable user embeddings with transparent, natural language profiles that holistically describe a user's preferences. We show that instruction prompting can be used with large language models to create fluent, informative, and personalized user profiles based on user reviews. By encoding user preferences using natural language, we are able to develop a recommender system that is both transparent and scrutable. Experimental results show that our method has comparable performance to popular recommender models. In addition, we show that by adding new preferences to the NL profile, we can quickly provide updated recommendations. For future work, we plan to investigate multi-turn updates to NL profiles to simulate user updates over time and explore how scrutability can be used to adjust recommendations to be less biased.


\section{Limitations}
\label{sec:ethical_considerations}
The performance of UPR is heavily dependent on the quality of the user reviews in the dataset, meaning that preferences that are not explicitly mentioned often in the review corpora will not be captured in the NL profile. Furthermore, the average rating in both datasets is 4.0/5.0, meaning that the vast majority of NL profiles contain \textit{only} positive preferences. Thus, the model does not have many negative preferences to train on, making it difficult to edit a negative preferences to the profile (shown in Appendix~\ref{subsec:appendix-negative-pref}).

Another major limitation of using LLMs for recommendation is that they are notably slower than traditional recommender systems. Thus, users would need to wait a significantly longer amount of time to receive recommendations, which can lead to lower levels of satisfaction.

Finally, the model is bottlenecked by the information contained in the NL profile, meaning that the length of the profile and the number of features impacts recommendation performance. However, there is a concern that a verbose NL profile increases cognitive load for users. In terms of practical utility, making the NL profile too long and difficult to scrutinize negatively impacts usability for users. Future work can explore the tradeoffs between recommendation performance and human preferences.

\section{Ethical Considerations}
There are several ethical considerations to consider when using large language models for recommendation. Firstly, LLMs may show bias towards more popular items, which can lead to lesser-known items being ignored. Since recommender systems learn from user interactions, a bias amplification loop can be created that causes a greater discrepancy between popular and less-popular items~\cite{ChenBias2023}. Furthermore, LLMs have also been criticized for unfairness with regards to underrepresented demographics, highlighting the importance of metrics that evaluate various social biases in LLMs~\cite{salutari2023quantifying}. Finally, hallucinations are a problematic issue in LLMs because the model may generate plausible, but incorrect information. In the case of recommender systems, the model might create an NL profile with incorrect information or recommend an item that does not exist~\cite{li2023large}. Consequently, safeguards must be added to ensure that any output from the language model is truthful and accurate.
\bibliography{sample-base}

\begin{thebibliography}{38}
\expandafter\ifx\csname natexlab\endcsname\relax\def\natexlab#1{#1}\fi

\bibitem[{Aggarwal(2016)}]{books/sp/Aggarwal16}
Charu~C. Aggarwal. 2016.
\newblock \emph{Recommender Systems - The Textbook}.
\newblock Springer.

\bibitem[{Balog et~al.(2019)Balog, Radlinski, and Arakelyan}]{transparent_rec}
Krisztian Balog, Filip Radlinski, and Shushan Arakelyan. 2019.
\newblock \href {https://dl.acm.org/citation.cfm?id=3331211} {Transparent, scrutable and explainable user models for personalized recommendation}.
\newblock In \emph{Proceedings of the 42nd International ACM SIGIR Conference on Research and Development in Information Retrieval (SIGIR '19)}.

\bibitem[{Bird et~al.(2009)Bird, Klein, and Loper}]{nltk}
Steven Bird, Ewan Klein, and Edward Loper. 2009.
\newblock \emph{Natural Language Processing with Python}, 1st edition.
\newblock O'Reilly Media, Inc.

\bibitem[{Bogers(2018)}]{DBLP:series/lncs/Bogers18}
Toine Bogers. 2018.
\newblock \href {https://doi.org/10.1007/978-3-319-90092-6\_12} {Tag-based recommendation}.
\newblock In Peter Brusilovsky and Daqing He, editors, \emph{Social Information Access - Systems and Technologies}, volume 10100 of \emph{Lecture Notes in Computer Science}, pages 441--479. Springer.

\bibitem[{Brown et~al.(2020)Brown, Mann, Ryder, Subbiah, Kaplan, Dhariwal, Neelakantan, Shyam, Sastry, Askell et~al.}]{brown2020language}
Tom Brown, Benjamin Mann, Nick Ryder, Melanie Subbiah, Jared~D Kaplan, Prafulla Dhariwal, Arvind Neelakantan, Pranav Shyam, Girish Sastry, Amanda Askell, et~al. 2020.
\newblock Language models are few-shot learners.
\newblock \emph{Advances in neural information processing systems}, 33:1877--1901.

\bibitem[{Chang et~al.(2015)Chang, Harper, and Terveen}]{set_preferences}
Shuo Chang, F.~Maxwell Harper, and Loren Terveen. 2015.
\newblock \href {https://doi.org/10.1145/2675133.2675210} {Using groups of items for preference elicitation in recommender systems}.
\newblock In \emph{Proceedings of the 18th ACM Conference on Computer Supported Cooperative Work \& Social Computing}, CSCW '15, page 1258–1269, New York, NY, USA. Association for Computing Machinery.

\bibitem[{Chen et~al.(2023)Chen, Dong, Wang, Feng, Wang, and He}]{ChenBias2023}
Jiawei Chen, Hande Dong, Xiang Wang, Fuli Feng, Meng Wang, and Xiangnan He. 2023.
\newblock \href {https://doi.org/10.1145/3564284} {Bias and debias in recommender system: A survey and future directions}.
\newblock \emph{ACM Trans. Inf. Syst.}, 41(3).

\bibitem[{Chen et~al.(2020)Chen, Du, He, and Wang}]{Chen2020JIT2RAJ}
Xu~Chen, Changying Du, Xiuqiang He, and Jun Wang. 2020.
\newblock \href {https://api.semanticscholar.org/CorpusID:220730037} {Jit2r: A joint framework for item tagging and tag-based recommendation}.
\newblock \emph{Proceedings of the 43rd International ACM SIGIR Conference on Research and Development in Information Retrieval}.

\bibitem[{Chowdhery et~al.(2022)Chowdhery, Narang, Devlin, Bosma, Mishra, Roberts, Barham, Chung, Sutton, Gehrmann, Schuh, Shi, Tsvyashchenko, Maynez, Rao, Barnes, Tay, Shazeer, Prabhakaran, Reif, Du, Hutchinson, Pope, Bradbury, Austin, Isard, Gur-Ari, Yin, Duke, Levskaya, Ghemawat, Dev, Michalewski, Garcia, Misra, Robinson, Fedus, Zhou, Ippolito, Luan, Lim, Zoph, Spiridonov, Sepassi, Dohan, Agrawal, Omernick, Dai, Pillai, Pellat, Lewkowycz, Moreira, Child, Polozov, Lee, Zhou, Wang, Saeta, Diaz, Firat, Catasta, Wei, Meier-Hellstern, Eck, Dean, Petrov, and Fiedel}]{chowdhery2022palm}
Aakanksha Chowdhery, Sharan Narang, Jacob Devlin, Maarten Bosma, Gaurav Mishra, Adam Roberts, Paul Barham, Hyung~Won Chung, Charles Sutton, Sebastian Gehrmann, Parker Schuh, Kensen Shi, Sasha Tsvyashchenko, Joshua Maynez, Abhishek Rao, Parker Barnes, Yi~Tay, Noam Shazeer, Vinodkumar Prabhakaran, Emily Reif, Nan Du, Ben Hutchinson, Reiner Pope, James Bradbury, Jacob Austin, Michael Isard, Guy Gur-Ari, Pengcheng Yin, Toju Duke, Anselm Levskaya, Sanjay Ghemawat, Sunipa Dev, Henryk Michalewski, Xavier Garcia, Vedant Misra, Kevin Robinson, Liam Fedus, Denny Zhou, Daphne Ippolito, David Luan, Hyeontaek Lim, Barret Zoph, Alexander Spiridonov, Ryan Sepassi, David Dohan, Shivani Agrawal, Mark Omernick, Andrew~M. Dai, Thanumalayan~Sankaranarayana Pillai, Marie Pellat, Aitor Lewkowycz, Erica Moreira, Rewon Child, Oleksandr Polozov, Katherine Lee, Zongwei Zhou, Xuezhi Wang, Brennan Saeta, Mark Diaz, Orhan Firat, Michele Catasta, Jason Wei, Kathy Meier-Hellstern, Douglas Eck, Jeff Dean, Slav Petrov, and Noah Fiedel. 2022.
\newblock \href {http://arxiv.org/abs/2204.02311} {Palm: Scaling language modeling with pathways}.

\bibitem[{Deshpande and Karypis(2004)}]{item-knn}
Mukund Deshpande and George Karypis. 2004.
\newblock \href {https://doi.org/10.1145/963770.963776} {Item-based top-n recommendation algorithms}.
\newblock \emph{ACM Trans. Inf. Syst.}, 22(1):143–177.

\bibitem[{Geng et~al.(2022)Geng, Fu, Ge, Li, de~Melo, and Zhang}]{geng-etal-2022-improving}
Shijie Geng, Zuohui Fu, Yingqiang Ge, Lei Li, Gerard de~Melo, and Yongfeng Zhang. 2022.
\newblock \href {https://doi.org/10.18653/v1/2022.acl-long.20} {Improving personalized explanation generation through visualization}.
\newblock In \emph{Proceedings of the 60th Annual Meeting of the Association for Computational Linguistics (Volume 1: Long Papers)}, pages 244--255, Dublin, Ireland. Association for Computational Linguistics.

\bibitem[{Harper and Konstan(2015)}]{movielens}
F.~Maxwell Harper and Joseph~A. Konstan. 2015.
\newblock \href {https://doi.org/10.1145/2827872} {The movielens datasets: History and context}.
\newblock \emph{ACM Trans. Interact. Intell. Syst.}, 5(4).

\bibitem[{He and McAuley(2016)}]{He2016UpsAD}
Ruining He and Julian McAuley. 2016.
\newblock Ups and downs: Modeling the visual evolution of fashion trends with one-class collaborative filtering.
\newblock \emph{Proceedings of the 25th International Conference on World Wide Web}.

\bibitem[{He et~al.(2017)He, Liao, Zhang, Nie, Hu, and Chua}]{he2017neural}
Xiangnan He, Lizi Liao, Hanwang Zhang, Liqiang Nie, Xia Hu, and Tat-Seng Chua. 2017.
\newblock Neural collaborative filtering.
\newblock In \emph{Proceedings of the 26th international conference on world wide web}, pages 173--182.

\bibitem[{Jiang et~al.(2023)Jiang, Sablayrolles, Mensch, Bamford, Chaplot, de~las Casas, Bressand, Lengyel, Lample, Saulnier, Lavaud, Lachaux, Stock, Scao, Lavril, Wang, Lacroix, and Sayed}]{jiang2023mistral}
Albert~Q. Jiang, Alexandre Sablayrolles, Arthur Mensch, Chris Bamford, Devendra~Singh Chaplot, Diego de~las Casas, Florian Bressand, Gianna Lengyel, Guillaume Lample, Lucile Saulnier, Lélio~Renard Lavaud, Marie-Anne Lachaux, Pierre Stock, Teven~Le Scao, Thibaut Lavril, Thomas Wang, Timothée Lacroix, and William~El Sayed. 2023.
\newblock \href {http://arxiv.org/abs/2310.06825} {Mistral 7b}.

\bibitem[{Kingma and Ba(2015)}]{2015-kingma}
Diederik~P. Kingma and Jimmy Ba. 2015.
\newblock \href {http://dblp.uni-trier.de/db/conf/iclr/iclr2015.html#KingmaB14} {Adam: A method for stochastic optimization.}
\newblock In \emph{ICLR (Poster)}.

\bibitem[{Koren et~al.(2009)Koren, Bell, and Volinsky}]{2009korenmf}
Yehuda Koren, Robert Bell, and Chris Volinsky. 2009.
\newblock \href {https://doi.org/10.1109/MC.2009.263} {Matrix factorization techniques for recommender systems}.
\newblock \emph{Computer}, 42(8):30--37.

\bibitem[{Li et~al.(2020)Li, Zhang, and Chen}]{CIKM20-NETE}
Lei Li, Yongfeng Zhang, and Li~Chen. 2020.
\newblock Generate neural template explanations for recommendation.
\newblock In \emph{CIKM}.

\bibitem[{Li et~al.(2021)Li, Zhang, and Chen}]{li-etal-2021-personalized}
Lei Li, Yongfeng Zhang, and Li~Chen. 2021.
\newblock \href {https://doi.org/10.18653/v1/2021.acl-long.383} {Personalized transformer for explainable recommendation}.
\newblock In \emph{Proceedings of the 59th Annual Meeting of the Association for Computational Linguistics and the 11th International Joint Conference on Natural Language Processing (Volume 1: Long Papers)}, pages 4947--4957, Online. Association for Computational Linguistics.

\bibitem[{Li et~al.(2023{\natexlab{a}})Li, Zhang, and Chen}]{li2023personalized}
Lei Li, Yongfeng Zhang, and Li~Chen. 2023{\natexlab{a}}.
\newblock Personalized prompt learning for explainable recommendation.
\newblock \emph{ACM Transactions on Information Systems}, 41(4):1--26.

\bibitem[{Li et~al.(2023{\natexlab{b}})Li, Zhang, Liu, and Chen}]{li2023large}
Lei Li, Yongfeng Zhang, Dugang Liu, and Li~Chen. 2023{\natexlab{b}}.
\newblock Large language models for generative recommendation: A survey and visionary discussions.
\newblock \emph{arXiv preprint arXiv:2309.01157}.

\bibitem[{Mysore et~al.(2023{\natexlab{a}})Mysore, Jasim, Mccallum, and Zamani}]{mysoreEditableUserProfiles2023a}
Sheshera Mysore, Mahmood Jasim, Andrew Mccallum, and Hamed Zamani. 2023{\natexlab{a}}.
\newblock \href {https://doi.org/10.1145/3539618.3591677} {Editable {{User Profiles}} for {{Controllable Text Recommendations}}}.
\newblock In \emph{Proceedings of the 46th {{International ACM SIGIR Conference}} on {{Research}} and {{Development}} in {{Information Retrieval}}}, pages 993--1003, Taipei Taiwan. ACM.

\bibitem[{Mysore et~al.(2023{\natexlab{b}})Mysore, Mccallum, and Zamani}]{mysoreLargeLanguageModel2023}
Sheshera Mysore, Andrew Mccallum, and Hamed Zamani. 2023{\natexlab{b}}.
\newblock \href {https://doi.org/10.1145/3604915.3608829} {Large {{Language Model Augmented Narrative Driven Recommendations}}}.
\newblock In \emph{Proceedings of the 17th {{ACM Conference}} on {{Recommender Systems}}}, pages 777--783, Singapore Singapore. ACM.

\bibitem[{Radford et~al.(2019)Radford, Wu, Child, Luan, Amodei, and Sutskever}]{radford2019language}
Alec Radford, Jeff Wu, Rewon Child, David Luan, Dario Amodei, and Ilya Sutskever. 2019.
\newblock Language models are unsupervised multitask learners.

\bibitem[{Radlinski et~al.(2022)Radlinski, Balog, Diaz, Dixon, and Wedin}]{user_profile_transparent_rec}
Filip Radlinski, Krisztian Balog, Fernando Diaz, Lucas Dixon, and Ben Wedin. 2022.
\newblock \href {https://doi.org/10.1145/3477495.3531873} {On natural language user profiles for transparent and scrutable recommendation}.
\newblock In \emph{Proceedings of the 45th International ACM SIGIR Conference on Research and Development in Information Retrieval}, SIGIR '22, page 2863–2874, New York, NY, USA. Association for Computing Machinery.

\bibitem[{Sakai(2007)}]{sakai}
Tetsuya Sakai. 2007.
\newblock \href {https://doi.org/10.1145/1277741.1277756} {Alternatives to bpref}.
\newblock In \emph{Proceedings of the 30th Annual International ACM SIGIR Conference on Research and Development in Information Retrieval}, SIGIR '07, page 71–78, New York, NY, USA. Association for Computing Machinery.

\bibitem[{Salakhutdinov and Mnih(2007)}]{pmf}
Ruslan Salakhutdinov and Andriy Mnih. 2007.
\newblock Probabilistic matrix factorization.
\newblock In \emph{Proceedings of the 20th International Conference on Neural Information Processing Systems}, NIPS'07, page 1257–1264, Red Hook, NY, USA. Curran Associates Inc.

\bibitem[{Salutari et~al.(2023)Salutari, Ramos, Rahmani, Linguaglossa, and Lipani}]{salutari2023quantifying}
Flavia Salutari, Jerome Ramos, Hossein~A Rahmani, Leonardo Linguaglossa, and Aldo Lipani. 2023.
\newblock Quantifying the bias of transformer-based language models for african american english in masked language modeling.
\newblock In \emph{Pacific-Asia Conference on Knowledge Discovery and Data Mining}, pages 532--543. Springer Nature Switzerland Cham.

\bibitem[{Sanner et~al.(2023)Sanner, Balog, Radlinski, Wedin, and Dixon}]{sanner2023large}
Scott Sanner, Krisztian Balog, Filip Radlinski, Ben Wedin, and Lucas Dixon. 2023.
\newblock Large language models are competitive near cold-start recommenders for language-and item-based preferences.
\newblock In \emph{Proceedings of the 17th ACM Conference on Recommender Systems}, pages 890--896.

\bibitem[{Sen et~al.(2007)Sen, Harper, LaPitz, and Riedl}]{tag_quality}
Shilad Sen, F.~Maxwell Harper, Adam LaPitz, and John Riedl. 2007.
\newblock \href {https://doi.org/10.1145/1316624.1316678} {The quest for quality tags}.
\newblock In \emph{Proceedings of the 2007 ACM International Conference on Supporting Group Work}, GROUP '07, page 361–370, New York, NY, USA. Association for Computing Machinery.

\bibitem[{Sharma et~al.(2019)Sharma, Harper, and Karypis}]{Sharma_2019}
Mohit Sharma, F.~Maxwell Harper, and George Karypis. 2019.
\newblock \href {https://doi.org/10.1145/3326128} {Learning from sets of items in recommender systems}.
\newblock \emph{ACM Transactions on Interactive Intelligent Systems}, 9(4):1–26.

\bibitem[{Tintarev and Masthoff(2015)}]{tintarev_explain_rec}
Nava Tintarev and Judith Masthoff. 2015.
\newblock \href {https://doi.org/10.1007/978-1-4899-7637-6_10} {\emph{Explaining Recommendations: Design and Evaluation}}, 2 edition, pages 353--382. Springer US.

\bibitem[{Touvron et~al.(2023)Touvron, Martin, Stone, Albert, Almahairi, Babaei, Bashlykov, Batra, Bhargava, Bhosale et~al.}]{touvron2023llama}
Hugo Touvron, Louis Martin, Kevin Stone, Peter Albert, Amjad Almahairi, Yasmine Babaei, Nikolay Bashlykov, Soumya Batra, Prajjwal Bhargava, Shruti Bhosale, et~al. 2023.
\newblock Llama 2: Open foundation and fine-tuned chat models.
\newblock \emph{arXiv preprint arXiv:2307.09288}.

\bibitem[{Tran et~al.(2021)Tran, Ghazimatin, and Saha~Roy}]{counterfactual_hiep_2021}
Khanh~Hiep Tran, Azin Ghazimatin, and Rishiraj Saha~Roy. 2021.
\newblock \href {https://doi.org/10.1145/3404835.3463005} {Counterfactual explanations for neural recommenders}.
\newblock In \emph{Proceedings of the 44th International ACM SIGIR Conference on Research and Development in Information Retrieval}, SIGIR '21, page 1627–1631, New York, NY, USA. Association for Computing Machinery.

\bibitem[{Truong et~al.(2021)Truong, Salah, and Lauw}]{truong2021multi}
Quoc-Tuan Truong, Aghiles Salah, and Hady Lauw. 2021.
\newblock Multi-modal recommender systems: Hands-on exploration.
\newblock In \emph{Fifteenth ACM Conference on Recommender Systems}, pages 834--837.

\bibitem[{Xie et~al.(2022)Xie, Singh, McAuley, and Majumder}]{Xie2022FactualAI}
Zhouhang Xie, Sameer Singh, Julian McAuley, and Bodhisattwa~Prasad Majumder. 2022.
\newblock Factual and informative review generation for explainable recommendation.
\newblock \emph{ArXiv}, abs/2209.12613.

\bibitem[{Zhang and Chen(2020)}]{explainable_rec}
Yongfeng Zhang and Xu~Chen. 2020.
\newblock \href {https://doi.org/10.1561/1500000066} {Explainable recommendation: A survey and new perspectives}.
\newblock \emph{Found. Trends Inf. Retr.}, 14(1):1–101.

\bibitem[{Zhang et~al.(2014)Zhang, Zhang, Zhang, Liu, and Ma}]{SIGIR14-Sentires}
Yongfeng Zhang, Haochen Zhang, Min Zhang, Yiqun Liu, and Shaoping Ma. 2014.
\newblock Do users rate or review? boost phrase-level sentiment labeling with review-level sentiment classification.
\newblock In \emph{SIGIR}.

\end{thebibliography}
\bibliographystyle{acl_natbib}

\appendix

\section{Appendix}
\label{sec:appendix}
\subsection{NL Profile Generation}
\label{subsec:appendix-prof-gen}
To generate the profiles, we use the top 5 features calculated per user in Section~\ref{subsec:rank_pref}. We then use 5 random reviews from each feature as input for the prompt, shown in Table~\ref{tab:example-prompts}. We set the max tokens generated to 200, temperature to 0.7, and seed to 0.

\subsection{Recommendation Task}
\label{subsec:appendix-rec-task}
We encode and fine-tune our model using the following prompt:\\

\noindent``\{\textbf{profile}\} Based on my user profile, from a scale of 1 to 5 (1 being the lowest and 5 being the highest), i would give \{\textbf{title}\} a rating of''\\

where \{\textbf{profile}\} is the user profile and \{\textbf{title}\} is the target item. In order to improve performance, we scale ratings from 1-5 to 0-1 when updating the loss per batch. Ratings are scaled back to their original values during evaluation. 
\subsection{Implementation Details}
For the recommendation task, we experiment with the number of features $k$ set between 1 to 5. For hyperparameter tuning, we experiment with a learning rate of 1e-3, 3e-4, 1e-5, learning scheduler of linear and cosine, and batch sizes of 8, 16, and 32. In addition we train the model for 10 epochs, with early stopping after 3 epochs of no improvement.

\subsection{Qualitative Test Study}
The participants are master's students from the United Kingdom who volunteered to participate in the study. The study took about 30 minutes on average. All participants agreed to allow the data to reported for research purposes only. There is no identifiable or sensitive information recorded in the user study.

Before annotating the samples, the participants were provided with both positive and negative examples for each of the four questions to help them better understand the task. The participants were then shown every sample from the randomly selected pool of profiles and were asked to indicate whether the profile fulfilled each of the four criteria.

\subsection{Adding New Preferences to NL Profile}
\label{subsec:appendix-editing-prof}
When sampling users and items from the test set, we use seeds=$\{0, 42, 100, 200, 300\}$. To generate the edited profile set $C$, we a few shot prompt, shown in Table~\ref{tab:few-shot-edit}, with max token set to 300, temperature of 0.7, and seed of 0.

During experimentation, we found that we needed to increase the max token size of the profiles to yield a change in recommendations because the profile needs more tokens to express a new preference in the NL profile. To obtain recommendations, we use the same prompt in Appendix~\ref{subsec:appendix-rec-task} using our fine-tuned model at inference time.

\begin{figure}[!t]
    \centering
    \includegraphics[width=\linewidth]{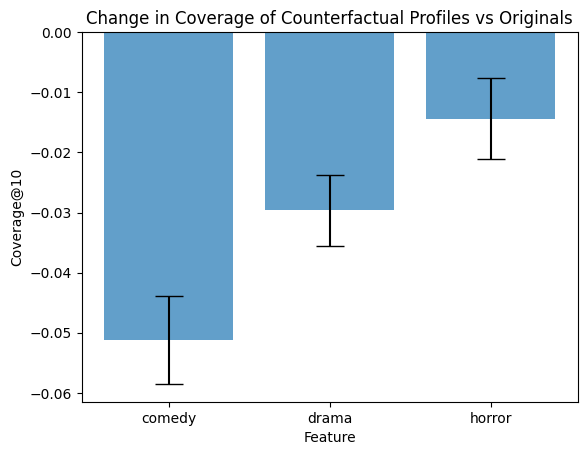}
    \caption{Change in Coverage@10 between counterfactual profile versus original profiles for a target feature.}
    \label{fig:negative-coverage}
\end{figure}

\subsection{Editing Negative Preferences to NL Profile}
\label{subsec:appendix-negative-pref}
We experiment with editing an existing positive preference towards target feature $t$ in a NL profile to a negative preference to see if a user can receive less recommendations of $t$ by scrutinizing the profile. We repeat the same sampling method in Section~\ref{subsec:new_preferences}, but instead select 200 users that positively mention target feature $t$ in their profile. We then use the following prompt with Llama2 to counterfactually edit the profile such that the user does not like $t$.\\

\noindent "Rewrite the user profile so that the user does not like beach movies. Keep the profile as similar as possible for all other preferences: \{\textbf{profile}\}."\\

To keep the counterfactual profiles as similar as possible to the original profiles, excluding $t$, we use the same hyperparameters in Appendix~\ref{subsec:appendix-prof-gen}.

We record the change in coverage for the top-10 recommendations between the counterfactual versus original profiles in Figure~\ref{fig:negative-coverage}. We observe that the coverage drops, as expected, but the overall change is minor across the three features we tested. This can be attributed to the strong skew towards positive reviews in the data, meaning that vast majority of NL profiles used during training do not contain any negative preferences. In future work, we plan to train the model using more negative preferences in order to realize dislikes in the updated recommendations.

\begin{table*}[]
    \centering
    \caption{We prompt an instruction-tuned LLM with the relevant reviews to generate cohesive and relevant NL profiles. The result is one NL profile included in our dataset.}
    \label{tab:example-prompts}
    \begin{tabular}{p{12cm}}
    \hline
    \textbf{Instruction}: Summarize in a single paragraph using the first person my general movie and tv preferences based on my reviews. Do not mention the word reviews.\\
    \textbf{Reviews}: \\
    - not the most brilliant actress even in the best of circumstances\\
    - in the past 5 years or so i've watched close to 8 to 10 films about elizabeth i\\
    ...\\
    - bergman \& fellini ( tim burton in this century ) remain my all-time favorite directors \\
    \hline
    \textbf{Result}: \\
      I enjoy watching movies and TV shows, particularly those with strong performances and well-crafted stories. I have a preference for classic films and directors such as Bergman and Fellini, and I appreciate a good adaptation of a play. I also enjoy watching actors take on complex roles and explore their craft. While I have mixed opinions on some actresses, I generally enjoy watching a talented performance. Recently, I have been drawn to films about Elizabeth I, and I appreciate a good historical drama. Additionally, I find it refreshing when a director can successfully weave together multiple storylines and timelines into a cohesive and engaging film.\\
      \hline
    \end{tabular}
    \label{tab:my_label}
\end{table*}

\begin{table*}[]
    \centering
    \caption{A few shot prompt example to add a new preference towards target feature $t$ to the profile. For all few shot prompts used for this experiment, refer to our repository.}
    \label{tab:example-prompts2}
    \begin{tabular}{p{12cm}}
    \hline
    Modify the user profile so that the user also likes comedy movies. Keep all the profile as similar as possible for all other preferences. \\

\textbf{Original Profile}: I generally prefer action-packed movies and TV shows with plenty of stunts and excitement, but I also appreciate subtle moments of quiet subtlety and powerful performances. I prefer to watch DVDs instead of most TV shows, and I have a special interest in the works of certain directors like Steven Spielberg and Michael Mann. I have been impressed by the recent works of directors like Steven Soderbergh, and I believe they are making great strides in modern cinema. \\
\textbf{New Profile}: I generally prefer action-packed movies and TV shows with plenty of stunts and excitement, and now I've found a liking for comedy movies too, enjoying their humor alongside the subtle moments of quiet subtlety and powerful performances that I appreciate. I prefer to watch DVDs over TV shows, maintaining a special interest in directors like Steven Spielberg and Michael Mann. The recent works of directors like Steven Soderbergh also impress me, as I believe they contribute significantly to modern cinema.\\

\textbf{Original Profile}: As for my movie preferences, I tend to enjoy films that offer a unique perspective and don't rely too heavily on cliches. I appreciate when there is a clear focus on character development and backstory, but I also enjoy when filmmakers take risks and try something new. I have mixed feelings about movies that try to tackle complex issues like bioterrorism, as they can often feel overly sensationalized or heavy-handed. Ultimately, I'm drawn to movies that offer a fresh take on familiar themes and have a strong sense of style and pacing.\\
\textbf{New Profile}: As for my movie preferences, I tend to enjoy films that offer a unique perspective, including comedy movies, and steer clear of relying too heavily on cliches. I appreciate clear focus on character development and backstory, and value when filmmakers take risks and introduce humor alongside their innovation. My views on movies tackling complex issues like bioterrorism are mixed, as they can feel sensationalized. Ultimately, I'm drawn to movies with a fresh take on familiar themes, strong sense of style, pacing, and those that blend genres effectively, including smart comedies that offer insightful laughs. \\
\textbf{Original Profile}: I have a preference for movies and TV shows that showcase talented actors, particularly those who can bring depth and nuance to their roles. While I appreciate the performances of actors like John Malkovich and Danny DeVito, I sometimes wish for a more convincing fit in certain roles. However, I also enjoy a good story and will often watch movies and shows regardless of the actors involved. I have a tendency to be critical in my reviews, but I also believe in being honest and providing constructive feedback. \\

\hline

\textbf{Result}: \\
  I have a preference for movies and TV shows that showcase talented actors, including those who excel in comedy, and appreciate depth and nuance in their roles. While I admire the performances of actors like John Malkovich and Danny DeVito, I sometimes wish for a more convincing fit in certain roles. However, I also enjoy a good story and will often watch movies and shows regardless of the actors involved. I strive to be honest and constructive in my reviews, providing valuable feedback for the benefit of both creators and fellow viewers.\\
  \hline
    \end{tabular}
    \label{tab:few-shot-edit}
\end{table*}


\end{document}